\documentclass[newstyle,twocolumn,proceedings]{rmaa}

\usepackage{rmaacite}

\renewcommand{\P}[1]{%
\ifnum#1=1\hbox{OW~168--326E}\fi
\ifnum#1=2\hbox{OW~167--317}\fi
\ifnum#1=3\hbox{OW~163--317}\fi
\ifnum#1=5\hbox{OW~158--323}\fi
\ifnum#1=0\hbox{OW~171--334}\fi}
\makeatletter

\title{Statistics and multiwavelength synthesis models: towards a new
  generation of synthesis models}
\author{M. Cervi\~no\altaffilmark{1,2}
  \altaffiltext{1}{LAEFF (INTA).}
  \altaffiltext{2}{IAA (CSIC).}}

\fulladdresses{
\item M. Cervi\~no: LAEFF (INTA), Apdo. 50727, Madrid 28080, Spain
(mcs@laeff.esa.es)
\item and IAA (CSIC), Camino Bajo de Hu\'etor 24, Granada 18080, Spain}

\shortauthor{Cervi\~no}
\shorttitle{Towards a new generation of Synthesis models}


\abstract{
In this contribution I present my current work in a new generation of
evolutionary synthesis models that compute the multiwavelength energy
distribution (from gamma-rays to radio) as well as the associated
dispersion for young stellar systems. I will also show some statistical
effects that may appear in the analysis of surveys, like bimodal or
multi-modal distributions and bias when color indices computed by the codes
are compared with observations.  Such new generation of synthesis models
may be useful for the analysis of the data expected from GTC.}

\resumen{ 
En esta contribuci\'on presento mi trabajo en el desarrollo de una nueva
generaci\'on de modelos de s\'\i ntesis evolutiva que calculan tanto el
espectro multirango (desde rayos-$\gamma$ a ondas de radio) como la
dispersi\'on asociada en c\'umulos estelares j\'ovenes. Tambi\'en muestro
algunos efectos estad\'\i sticos que pueden aparecer en en an\'alisis de
conjuntos de datos, como distribuciones bimodales o multimodales y sesgos
cuando se comparan los resultados de los modelos de s\'\i ntesis con datos
reales. Esta nueva generaci\'on de modelos de s\'\i ntesis puede ser una
herramienta muy \'util para el an\'alisis de los datos que se esperan
obtener con el GTC.}

\listofauthors{M.~Cervi\~no}
\indexauthor{M.~Cervi\~no}

\begin{document}

\maketitle


\section{Introduction}

Taking into account the discreteness of the stellar population, the
predictions of any model that relies on an Initial Mass Function (IMF) are
only exact {\it under the assumption of an infinite number of
stars}. Otherwise, they only give a {\it mean value} of a probability
distribution. The relevance of such fluctuations in the results of
synthesis models is obvious in the case of massive stars and young
clusters, {\it but they also affect the models of older clusters dominated
by the emission of low-mass stars} since small variations in the initial
mass/number of stars in a given mass range, can produce different numbers
of, e.g., AGB stars at a given age, which in turn produce large variations
in the resulting observable.

So, for the comparison of models with observational data it is necessary to
obtain not only the mean value of the observable, but also, at least, the
corresponding dispersion of the computed observable due to the
discreteness of the stellar population, and, ideally, the underling
probability density distributions. The dispersion due to a finite number of
stars in real systems can be evaluated theoretically, as it has been shown
in Cervi\~no et al. (2002).

\section{Some examples: multiwavelength emission and color-color diagrams}

\begin{figure*}
\centerline{\includegraphics[width=11pc,angle=270]{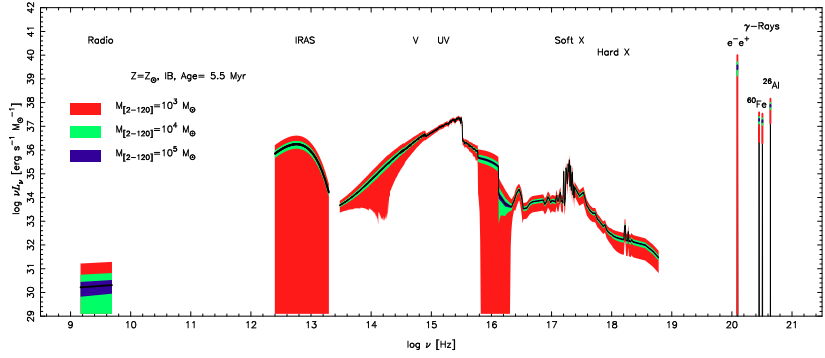}}
\caption{Analytical 90\% Confidence Level for the multiwavelength spectrum
for a 5.5 Myr old burst.}
\label{fig:multi}
\end{figure*}

In Fig. \ref{fig:multi} I show the 90\% confidence level for the
multiwavelength spectrum for a 5.5 Myr old burst obtained analytically from
our code. Examples of statistical effects of a large number of Monte Carlo
simulations and comparisons with the dispersion obtained theoretically for
different observables can be found in Cervi\~no et al. (2000, 2001, 2002).

\begin{figure}
\centerline{\includegraphics[width=6cm,angle=270]{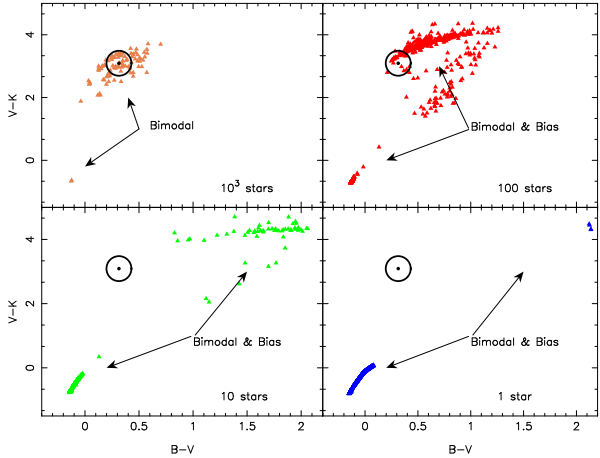}}
\caption{Monte Carlo simulations of the B--V vs. V--K colors of 10 Myr old
stellar clusters with 1000, 100, 10 and 1 stars in the mass range 2--120
M$_\odot$ following a Salpeter IMF slope (triangles). The theoretical
result is showed with the $\odot$ symbol.}
\label{fig:color}
\end{figure}

In Fig. \ref{fig:color} I show the B--V vs. V--K colors for a 10 Myr old
star forming region. Note that {\it it is assumed that colors are
independent on the number of stars in the cluster}, so I have test this
hypothesis with 10$^3$ Monte Carlo simulation of clusters (triangles) with
1000, 100, 10 and 1 stars in the mass range 2--120 M$_\odot$ following a
Salpeter IMF slope. The theoretical result is shown with the $\odot$
symbol.  The Figure shows two important effects: {\bf (a)} There are
bimodal (multi-modal) distributions of the colors. {\bf (b)} The mean
value obtained by infinite populated IMF (analytical result) has a bias if
it is used for the analysis clusters with a small number of stars.  As
reference, the mean mass of a cluster with 10$^3$ stars in the given mass
limits is $\approx 6 \times 10^3$ M$_\odot$, it means $2 \times 10^4$
M$_\odot$ if the value of lower mass limit of the IMF is 0.08 M$_\odot$.
Both, bimodality and bias, are a natural effect of statistics when only a
few stars dominate the observable.  Note that it is dependent on the
observable: {\it Example 1:} A cluster with 10$^5$ stars in the mass range
2 -- 120 M$_\odot$ with a Salpeter IMF slope have about only 10 WR stars!.
{\it Example 2:} The effect on the colors of globular clusters (dominated
by a few AGB star) can be relevant (see Bruzual 2001 and references therein
for details). Additionally, these bimodality effects produce a bias when
ratios and logarithm quantities computed by synthesis codes are compared
with observed data (but such a bias is not present for quantities like
luminosities even for ``clusters'' with only 1 star). This subject is
addressed in Cervi\~no \& Valls-Gabaud (submitted).

\adjustfinalcols

\section{Discussion}

Statistical effects may be specially relevant {\it (i)} In the analysis of
individual systems. In particular for the galaxy building blocks if they
are small clusters. {\it (ii)} In the analysis of the integrated spectra
obtained with high-resolution instruments when not all the region is
covered by the slit.  {\it (iii)} In chemical evolutionary models
(Cervi\~no \& Moll\'a, submitted), as far as they are dependent on the
yields production form Supernovae that are intrinsically ``rare'' events.
{\it (iv)} In the analysis of surveys, in special for the case of galaxies
with emission lines produced by massive stars, that are intrinsically a
small fraction of the total number of stars but dominate the emission....
In all these cases, the use of multiwavelength observations and
self-consistent synthesis models that includes such statistical effects and
computes the correlation between different wavelengths will be an asset to
constrain safely the physical properties of stellar systems (form small
star forming regions to high-redshift galaxies).  The current model outputs
can be found in {\tt http://www.laeff.esa.es/mcs/SED/}.





\begin{thebibliography}
\bibitem[Bruzual<2001>]{Bru01b}
  Bruzual, G. 2001 in {\it Extragalactic Star Clusters}, 
  IAU Symp. 207, E.K. Grebel, D. Geisler \& D. Minniti (eds.), in press
\bibitem[Cervi\~no et al.<2000>]{CLC00}
  Cervi\~no, M., Luridiana, V., \& Castander, F.J. 2000, A\&A, 360, L5
\bibitem[Cervi\~no et al.<2001>]{Cetal01a}
   Cervi\~no, M., G\'omez-Flechoso, M.A., Castander, F.J., Schaerer, D.,  
   Moll\'a, M., Kn\"odlseder, J., \& Luridiana, V. 2001, 
   A\&A, 376, 422
\bibitem[Cervi\~no et al.<2002>]{Cetal01b}
  Cervi\~no, M., Valls-Gabaud, D., Luridiana, V. \& Mas-Hesse, J.M. 2002, 
  A\&A, 381, 51 
\end{thebibliography}
\end{document}